\documentstyle[11pt,aaspp4]{article}
\def\xxi0{\hbox{${\xi^\prime}_0$}}
\def\th0{\hbox{$\theta_0$}}
\def\r0{\hbox{$\varpi_0$}}

\def\f0{\hbox{$\psi_0$}}

\def\v1{\hbox{$v_{K1}$}}
\def\ro1{\hbox{$\rho_1$}}
\def\n1{\hbox{$n_1$}}
\def\b1{\hbox{$B_1$}}

\def\mstar{\hbox{$M_\ast$}}
\def\rstar{\hbox{$R_\ast$}}
\def\tstar{\hbox{$T_\ast$}}

\def\msol{\hbox{${\rm M}_\odot $}}
\def\rsol{\hbox{${\rm R}_\odot $}}

\def\kelvin{{\rm K}}
\def\gauss{{\rm G}}

\def\kb{\hbox{$k_{\rm B}$}}

\def\subsectn#1{
    \addtocounter{subsection}{1}
    \setcounter{subsubsection}{0}
    {\ifnum\value{subsection}>1 \vskip 0.25 truein \penalty 10000 \fi}
    \penalty 10000
    {\centerline {\it \thesubsection.\ #1}}
    \penalty 10000      
    \bigskip
    \penalty 10000
    \index{#1}
    }
\def\subsubsectn#1{
    \addtocounter{subsubsection}{1}
    {\ifnum\value{subsubsection}>1 \vskip 0.125 truein \penalty 10000 \fi}
    \penalty 10000
    {\centerline {\it \thesubsubsection.\ #1}}
    \penalty 10000
    \bigskip
    \penalty 10000
    \index{#1}
    }
\newcommand{\ag}{\mbox{ \raisebox{-.4ex}{$\stackrel{\textstyle >}{\sim}$} }}

\def\pccm{\mbox{$\rm cm^{-3}$}}

\def\kev{\mbox{\rm keV}}
\def\lx{\mbox{$L_{ x}$}}
\def\lradio{\mbox{$L_{ r}$}}

\def\gstar{\mbox{$g_\ast$}}
\def\betastar{\mbox{$\beta_\ast$}}
\def\rco{\mbox{$r_{\rm co}$}}
\def\omegastar{\mbox{$\Omega_\ast$}}
\def\omegakepler{\mbox{$\Omega_{ k}$}}
\def\vkepler{\mbox{$v_{ k}$}}
\def\ropen{\mbox{$r_{ t}$}}

\def\ralfven{\mbox{$r_{ a}$}}
\def\rslow{\mbox{$r_{ s}$}}
\def\rfast{\mbox{$r_{ f}$}}
\def\tff{\mbox{$t_{\rm ff}$}}

\def\tins{\mbox{$t_{\rm ins}$}}
\def\tshear{\mbox{$t_\varphi$}}

\begin{document}

\title{A Critique of Current Magnetic-Accretion \\
Models for Classical T Tauri Stars}
\author{Pedro N. Safier}
\affil{University of Maryland,\\
 Laboratory for Millimeter-Wave Astronomy,\\
 College Park, MD 20742;\\ e-mail: {\it
    safier@astro.umd.edu\/}}
%
\begin{abstract}
%
%
Current magnetic-accretion models for classical T-Tauri stars rely on a
{\it strong}, {\it dipolar\/} magnetic field of stellar origin to funnel
the disk material onto the star, and assume a steady-state.
In this paper, I critically examine the physical basis of these
models in light of the observational evidence
and our
knowledge of magnetic fields in low-mass stars, and find it lacking. 
 I also argue that magnetic accretion onto these stars
is inherently a time-dependent problem, and that a steady-state is not
warranted.

Finally, directions for future work towards fully-consistent models
are pointed out.
\end{abstract}

\section{Introduction}
%
%
It is now widely accepted that classical T Tauri stars  are accreting
material from their circumstellar disks. So far, this is the most successful
explanation for the large fluxes in the blue and UV regions of
the spectrum that characterize these stars (Haro \& Herbig
1954\markcite{hh54}; Walker, 1956\markcite{walk56}), and the filling-in
or ``veiling'' of the absorption spectrum longward of $3800\,$\AA\
(Joy, 1949\markcite{joy49}; Haro \& Herbig, 1954\markcite{hh54}). 
The strongest evidence for accretion is the detection of
redshifted absorption in upper Balmer and permitted
metallic lines---at velocities of up to
several hundred ${\rm km\,s^{-1}}$---in T Tauri stars with large
optical and UV excesses 
(Walker, 1972\markcite{walk72}; Edwards et
al., 1994\markcite{edal94}, and references therein). 

Notwithstanding the observational evidence, our understanding of
the accretion process is sketchy---at best. Early attempts (Bertout,
Basri, \& Bouvier, 1988\markcite{bbb88}; Basri \& Bertout,
1989\markcite{bb89}), following the original suggestion by Lynden-Bell
\& Pringle (1974\markcite{lbp74}), modeled the accretion region
as  a shear boundary layer at the star-disk boundary. These empirical,
equatorial boundary layer models remained popular until the early 90's, even
though they are inconsistent with the detection of redshifted {\it
  absorption\/} lines.
Currently, the most popular picture
is that pioneered by K\"{o}nigl (1991\markcite{kon91}),
wherein the
disk material is funneled onto the star by a dipolar stellar magnetic
field. A variation of this scenario has been put forward
by Shu and collaborators (Shu et al.,
1994a,b\markcite{shu94a}\markcite{shu94b}; Ostriker \& Shu,
1995\markcite{oshu95}). A major appeal of these models is their
supposed ability to account for
the fact that classical T Tauri stars are slower rotators than their
naked siblings (see, e.g., Edwards et al., 1993,\markcite{edal93} and
references therein).

The purpose of this paper is to point out that the magnetic
topology assumed in 
these models is inconsistent with all the available data on T Tauri
stars (TTS) 
and our knowledge of magnetic fields in low-mass stars---a state of
affairs analogous to that in the late 80's, when boundary layer models
were widely popular, even though they were inconsistent with the 
spectroscopic evidence for non-equatorial mass infall. 

The paper is organized
as follows: in \S 2 I critically review the observational evidence that is
inconsistent with the current assumption of a strong stellar dipolar
field; the large-scale structure of a stellar field consistent with
the data, and its implications for the accretion process,
are examined in \S 3 and \S 4, respectively; and \S 5 contains my
conclusions. 
\section{The Observational Evidence}

Over the last two decades a wealth of data has accumulated which
indicates that the magnetic activity of 
T-Tauri stars (both classical and their naked siblings) is a scaled-up
version of that in low-mass, main-sequence stars, wherein 
the stellar magnetic field is spatially  intermittent and its topology
varies with time (see, e.g., Rosner, Golub, \& Vaiana, 1985; and
references therein).

The most dramatic evidence for dynamo-driven magnetic activity in
T-Tauri stars,
akin to that found
in late-type dwarfs, is the
unexpected discovery that T-Tauri stars---both classical (CTTS) and
naked (NTTS) T-Tauri stars---are strong emitters of X-rays
(with typical X-ray luminosities a factor $10^2$--$10^4$ above
solar levels), and that this emission is time-variable (for a
review, see Feigelson et al., 1991\markcite{fgv91}; and Montmerle et
al., 1993\markcite{montal93}). Recent observations of the Orion region
with {\it ASCA\/} (Yamauchi et al., 1996\markcite{yam96}) confirmed previous
findings that this
 emission can be attributed to a
multi-temperature plasma with at least two components: one at $\kb T
=0.7$--$1 \, \kev$ and a second at $\kb T \ag 2\,\kev$.

A clue to
the origin of the X-ray emission in
T-Tauri stars is that these objects extend the range of the
correlation between X-ray
luminosity, \lx, and
stellar equatorial velocity $v\sin i$  (Bouvier,
1990\markcite{bouv90}; Damiani \& Micela, 1995\markcite{dami95}) which
was found for low-mass dwarfs and other magnetically-active objects
(Rosner, Golub, \& Vaiana, 1985\markcite{ral85}).
 This so-called rotation-activity
connection supports the view that TTS stars have coronae
similar to those in late-type dwarfs, wherein spatially {\it
  intermittent \/}  magnetic fields with a {\rm time-variable \/} {\it
  topology\/} are responsible for the heating of the outer
atmosphere (Rosner, Golub, \& Vaiana, 1985\markcite{ral85}).

The strong variability of the X-ray emission in T-Tauri stars
(see Montmerle et al., 1983\markcite{montal83})
 lends further support to the idea that
the magnetic field in TTS is spatially intermittent and its topology
varies with time. X-ray flares in these objects are typically
$10^2$--$10^3$---and up to $10^4$---times stronger 
than the strongest solar flares  (Preibisch, Zinnecker,
\& Schmitt, 1993\markcite{pzs93}, and references therein). Note that
there is no free energy in a global dipolar field, and therefore such
field cannot drive a flaring corona (see, e.g., Priest,
1982;  Haisch, Strong, \& Rodon\`{o},
1991\markcite{hsr91}).

Additional evidence  for the dynamo-origin of the stellar magnetic field
in TTS is the detection at microwave wavelengths of non-thermal radio
emission from some of these objects (see, e.g., Chaing, Phillips, \&
Londsdale, 1996\markcite{cpl96}, and references therein; White,
Pallavicini, \& Kundu, 1992\markcite{wpk92}).
The luminosity of
the non-thermal radio emission in TTS, \lradio, extends to larger luminosities
 the correlation between \lx\
and \lradio\ found for low-mass dwarfs and other magnetically-active stars
(G\"{u}del \& Benz,1993\markcite{gube93}). This correlation, 
which extends over six
orders of magnitude in \lx\ and \lradio, suggests that the heating of
active coronae and particle acceleration are closely linked, as is the
case in solar flares. G\"{u}del \& Benz note that, according to the
relation between \lx\ and \lradio, the quiet sun stands out as a
radio-underluminous star. Therefore, stars that follow the \lx-\lradio\
relation---and in particular T-Tauri stars---must have a much higher
level of magnetic activity in their {\it quiescent\/} state than the quiet sun.

Irregular variability at all wavelengths is a distinguishing characteristic of
TTS (Joy, 1945\markcite{joy45}). However, it
is now well established that for many objects the optical light
curves have a periodic modulation (see, e.g., Herbst et al.,
1994\markcite{hhgw94}; Safier, 1995\markcite{saf95}, and references
therein) due to stellar spots that are colder than the surrounding
photosphere. These cold spots are regions of enhanced magnetic flux in
the photosphere (see, e.g., Gray, 1989\markcite{gray89}), and they
provide additional evidence that the magnetic field in TTS is
spatially intermittent.

Recent Doppler imaging of the NTTS
V410 Tau (Hatzes, 1995\markcite{hat95}, and references therein) and
HDE 283572 (Joncour, Bertout, \& Bouvier, 1994\markcite{jbb94})  confirms
the presence of cold stellar spots, but also show that their surface
distribution  on these stars is 
fundamentally different from that on the sun. In these NTTS, the
spots are
concentrated towards the polar caps---without any symmetry with
respect to the stellar equator---and the fractional coverage of
the stellar surface is very large.  This is consistent with the
results of Sch\"{u}ssler et al. (1996\markcite{sual96}) and DeLuca,
Fan, \& Saar (1997\markcite{dfs97}), who 
found that with increasing angular velocity and depth of the
convection zone, the Coriolis force dominates over magnetic bouyancy, and
the mean latitude of flux-emergence shifts towards the poles. 
The magnetically-active regions in the sun, where
X-ray flares originate, are located above
sunspots, and outside these regions the magnetic field is much weaker.
If the sun is any guidance, then these findings for NTTS
may be interpreted as suggesting that their magnetic fields near
the stellar equator are significantly weaker than those in the
cold spots. 

Finally, the failure to measure the surface fields of TTS by Zeeman
polarization experiments (Johnstone \& Penston, 1986\markcite{jp87},
1987\markcite{jp87}), and their detection by means of Zeeman
broadening observations (Basri, Marcy, \& Valenti,
1992\markcite{bmv92}),
 are consistent with a complex magnetic
topology. Recall that the Zeeman polarization depends on the orientation
of the magnetic field with respect to the observer; if the field
orientation changes along the line of sight, then the contributions  to the
total polarization from different magnetic field elements 
will nearly cancel out. On the other hand, the Zeeman-broadening
technique (Robinson, 1980\markcite{rob80}) relies on the measurement
of Zeeman-induced changes in the {\it unpolarized\/} line profiles.

It may seem paradoxical that T Tauri stars, which are fully convective
(Stahler, 1988\markcite{stah88}, and references therein), 
exhibit all the characteristics of dynamo-generated
magnetic fields,  when a variety of theoretical arguments 
suggest that the dynamo operative
in the sun---and late-type dwarfs with radiative cores---requires a strong
toroidal field anchored at the base of the convection zone (Parker,
1975\markcite{park75}; Golub et al.,1981\markcite{gol81}). 

However, Rosner (1980\markcite{ros80}) has
discussed how a magnetic dynamo could operate in a fully convective
star, and Durney, De  Young, \& Roxburgh (1993\markcite{ddr93}) first
showed in detail that such dynamo is possible.
Recent observations by Hodgkin,
Jameson, \& Steele (1995\markcite{hjs95}) 
show that
fully convective, young main-sequence 
stars have coronae, and suggest that the nature of
the stellar dynamo in these objects is different from that
in stars with radiative cores. 
In other words, although the
nature of the dynamo in fully convective stars is not fully
understood, the observations show that fully convective stars {\it
  do\/} have active coronae. Therefore, there is no reason why, in
principle, the magnetic fields in T Tauri stars could not be generated
by a stellar dynamo.

To summarize, the current observations of T-Tauri stars are consistent
with the notion that their magnetic field is generated by a stellar
dynamo. Dynamo fields cannot be axisymmetric (Cowling,
1934\markcite{cow34}), and, therefore, 
the magnetic field in these stars is most likely
spatially intermittent---with a variety of structures on different
spatial scales---and highly variable with time.
\section{The Large-Scale Structure of the Stellar Magnetic Field}

In current models of magnetic accretion, the region of interaction
between the stellar field and the circumstellar disk is located at
distances of order several stellar radii. It is usually argued that
the complexity of the stellar field is irrelevant because, at these
distances, the dipolar component of the field will dominate over
larger multipoles.
This is correct, but one may not conclude that the field has a closed
geometry at large distances from the star. The reason is that the
stellar magnetic field is embedded in hot, coronal gas---whose
presence is overlooked in current magnetic-accretion models---and the
thermal pressure eventually dominates over magnetic forces.
Once this happens, if the thermal pressure is large enough to overcome
the gravitational force, a stellar wind will flow (Parker,
1963\markcite{park63}). This stellar wind will drag and stretch the
magnetic field lines, changing the field topology into a configuration
wherein the field lines are mostly radial away from the star. Note
that, close to the star, regions with both open and closed field
lines will exist, but the large-scale structure of the field will be
predominantly open; Figure 1 illustrates this situation.

Does thermal pressure ever dominate over magnetic stresses in a
typical TTS corona? To answer this question, one has to find 
the ratio of the gas to magnetic pressure,
$\beta\equiv p_{\rm gas}/p_{\rm mag}$, as a function of distance from
the coronal base. To quantify matters, consider
the following example given by Rosner et al. (1995\markcite{ral95}). 
Assume---in light of the observed magnetic activity of
these stars---that the surface field can be regarded as a random and sparse
spatial
superposition of surface dipoles of fixed strength, with a
mean separation  $d$ between dipoles. At a height $z\equiv
r-\rstar$ above the stellar surface, the mean field-strength, $B$, from this
flux distribution is a function of $z$ and $d$ only, and 
$B(z)\propto d^3\,(z^2+d^2)^{-3/2}$. Hence, in
the far-field ($z\gg d$), $B(z) \propto (d/z)^3$
to first order in $d/z$. This means that the
magnetic field varies very rapidly with height as soon as one reaches
heights comparable to the typical correlation length of magnetic flux-bundles
at the stellar surface. 

Next, one has to find the radial dependence of the gas
pressure. In light of the strong dependence of the magnetic field
strength with distance, it suffices to use  an approximation for the
gas pressure that is appropriate for distances of order a few
stellar radii. Therefore, assume that the run of gas pressure with
distance is that for
a static,  isothermal atmosphere, where $p_{\rm gas}(r) \propto
\exp\left[\rstar(\rstar-r)/rH\right]$. Here $H\equiv 2\kb \tstar/m_{
  H}\,\gstar$ is the gravitational scale-height,
  \tstar\ is the---constant---temperature, and \gstar\ is the 
  gravitational acceleration at the stellar surface. The isothermal
  assumption is obviously an oversimplification, but it should be
  noted that, starting at the coronal base, the temperature will first
  increase with distance before it starts to drop. 

  In the sun, the gas
  temperature at the base of the corona is not uniform. The highest
  temperatures, $\tstar=2.5\times 10^6\,\kelvin$, 
  occur in active regions, wherein the magnetic field topology is
  mostly closed. These active regions appear in X-ray images as
  bright, closed loops. On the other hand, the lowest
  value of \tstar, $\tstar= 1.4$--$1.8\times10^6\,\kelvin$, is found in the
  so-called coronal holes, which are regions of open magnetic field
  lines where the bulk of the  solar wind originates\footnote{One of the most
  striking features of X-ray images of the solar corona is the
  presence of bright loops against a background of 
  diffuse emission and large, X-ray-dark regions---the so-called
  coronal holes.   However, one may not conclude that the hot
  gas ($T\ga 10^6\,\kelvin$) is confined to the coronal loops. The
  X-ray contrast between the loops and the coronal holes is due to the
  enhanced densities in the former (see, e.g., Munro \&
  Jackson,1977\markcite{mr77}; Priest, 1982).}.
  Bame et al. (1975\markcite{bame75}) deduced the gas temperature
  above coronal holes out to 
  distances $\la 4 \rsol$, and found it to be roughly constant.
  Therefore---and given the
  observational evidence for large amounts of hot coronal gas up to
  $T\sim 10^7\,\kelvin$ in TTS (see \S 2)---the isothermal assumption with
  $\tstar=10^6\,\kelvin$ should give an appropriate {\it lower\/}
  limit to the gas pressure out to $r\sim $few\,\rstar.

The distance $\ropen$
  where the gas pressure comes to dominate the
  magnetic pressure   is found by solving the equation $p_{\rm 
  gas}(\ropen) \sim B^2(\ropen)/8\pi$. In this example, the gas
  pressure will always overcome the magnetic pressure far enough from
  the star because $B^2 \propto r^{-6}$, whereas $p_{\rm gas}$
  approaches asymptotically a finite value. 
To be definite, I have plotted in Figure 2 the solution for \ropen\ as
  a function of 
  $d$ for different values of $B_\ast$ and $n_\ast$ for a 
  typical T-Tauri star with $\mstar=0.5\,\msol$ and
  $\rstar=2.5\,\rsol$, and using $\tstar=10^6\,\kelvin$.

  The values of $n_\ast$ and $B_\ast$ that I have chosen should be
  typical. At the base of the solar corona $n_\ast=10^8$--$10^9\pccm$, and
  because the transition 
  regions  between the chromospheres  and coronae of TTS are denser than
  in the sun (see, e.g.,
  Brown, Ferraz, \&  Jordan, 1984\markcite{bfj84}) these values of
  $n_\ast$ are   probably  lower limits. As for the value of $B_\ast$,
  Basri, Marcy, \& Valenti (1992) detected {\it photospheric\/} fields
  $\sim 1\,{\rm kG}$---a typical value in sunspots (see, e.g.,
  Priest, 1982).  Because the field strength inferred in active regions
  above sunspots is a factor $\sim 10$ smaller than the
  photospheric field (Priest, 1982), and, to repeat, because the {\it
  photospheric\/} field strengths measured in TTS are similar to those
  in the sun, a value $B_\ast=100\,\gauss$ is probably typical at the
  base of TTS's coronae. However, and for completeness, I have plotted
  in Figure 2 also the solution for $\ropen(d)$ for the unlikely case
  $B_\ast=10^3\,\gauss$. 

 The results in Figure 2 show that, for $d\la 0.1\,\rstar$ and $B_\ast
 \le 10^3\,\gauss$, $\ropen \la 3\,\rstar$. Therefore, for small
 mean-separations of the magnetic elements on the stellar surface, the
 thermal pressure dominates over the magnetic stresses beyond $\la
 3\,\rstar$ {\it right above\/} the closed structures. Between these
 loops, if the thermal pressure is large enough to drive a stellar
 wind, the radial field-lines will extend all the way to the coronal base
(see Figure 1). Even if $d\sim \rstar$, with $B_\ast\le 10^3\,\gauss$, thermal
 pressure will dominate most of the corona beyond $r \ll 3\,\rstar$,
 and the argument is as follows: because the number of magnetic
 elements that can be accomodated on the stellar surface with a mean
 separation $d$ scales as $d^{-2}$, if $d\sim \rstar$, only a {\it
 few\/} such elements will be 
 present. Right above these large loops, magnetic stresses may be
 dominant (depending on the value of $B_\ast$) beyond $\sim 3\,\rstar$;
 however, between these loops, again, the field lines will be radial all
 the way to  the coronal base if a thermally-driven wind is present
 (Figure 1), and a large fraction $\sim  (d/\rstar)^2\,\sim 1$ of the
 corona will be threaded by open 
 field-lines. Moreover, in light of the evidence for a predominance of
 high-latitude cold spots---as a result of the Coriolis force dominating over
 magnetic buoyancy (see \S 2)---most of these large, closed magnetic
 structures are likely to be away from the stellar equator, and will
 not intersect the disk's midplane\footnote{In light of the observed
 correlation between \lradio\ 
 and \lx, it is unlikely that most of the closed field-lines are in a
 few, very large loops. Though the emission measures derived from the
 hot component of the X-ray emission can be explained in terms of a
 few, dense, very large magnetic loops with material at temperatures
 $\sim 10^7\,\kelvin$, such isolated loops cannot account for the non-thermal
 electrons required to explain the observed non-thermal radio
 luminosities. On the other hand, a
 multitude of smaller loops undergoing frequent flaring can explain
 the observed X-ray luminosities at $T\sim 2\,\kev$, the non-thermal
 radio emission, {\it and\/}
 the correlation between \lradio\ and \lx\, (G\"{u}del,
 1997\markcite{gu97}).}.  

 In summary, the value $\ropen \la 3\,
 \rstar$ is a likely {\it upper\/} limit to the distance from a
 typical TTS where thermal pressure starts to dominate over magnetic
 stresses,  and the critical 
 question now is whether a thermally-driven stellar wind is possible. 

To answer this question,  I have
computed a Weber \& Davis (1967\markcite{wd67}) wind model using 
the 1-D generalization by Sakurai (1985\markcite{1985}), with
\mstar, \rstar, \tstar,  \betastar, and \omegastar\  as input
parameters, where $\betastar=8\pi n_\ast\,\kb\, T/B_\ast^2$ and
\omegastar\ is the stellar angular velocity. The Weber
\& Davis model (originally applied to the sun) assumes a steady-state,
polytropic flow confined to the equatorial 
plane of a rotating star; the generalization by Sakurai allows for
arbitrary rotation rates. Though this wind model strictly applies to
the stellar equatorial plane, it should 
be accurate enough for my purposes because of the small geometrical
thickness of accretion disks around TTS (see, e.g., Bertout, Basri, \&
Bouvier, 1988).

For the fiducial values $\mstar=0.5\,\msol$, $\rstar=2.5\,\rsol$,
 $\tstar=10^6\,\kelvin$, 
$\betastar=3.5\times 10^{-4}$ 
(corresponding to $B_\ast=100\,\gauss$ and $n_\ast=10^9\,\pccm$ at
$\tstar=10^6\,\kelvin$), and $\omegastar=10^{-5}\,{\rm s^{-1}}$---which
 gives a stellar rotation period of $7.3\,{\rm d}$, a typical value
 for CTTS (Bouvier et  al., 1993\markcite{bouv93})---I was able to
 find a wind solution that 
 becomes trans-sonic at
$\rslow=1.07\,\rstar$, trans-Alfv\'{e}nic at $\ralfven=1.59\,\rstar$,
and  trans-fast-magnetosonic at $\rfast=1.62\,\rstar$.

If one varies the values of $B_\ast$ and $n_\ast$, while holding the other
parameters constant, the values of \rslow, \ralfven, and \rfast\
change by only a few percent, with $\rfast - \ralfven \ll 1$. This
shows that these stellar winds are 
thermally driven---even though TTS are relatively fast rotators---and
therefore the stretching of the field lines is  very effective.

 If one chooses $\mstar=1\,\msol$ and
$\rstar=3\,\rsol$ (this value of \rstar\ corresponds to an age $\sim
0.5\,{\rm Myr}$ according to the evolutionary tracks of D'Antona \&
Mazzitelli [1994]\markcite{dm94}), then $\rslow=2.4\,\rstar$,
$\ralfven=2.8\,\rstar$, and $\rfast=3.6\,\rstar$;
 while for a $0.3\,\msol$ TTS with
$\rstar=2.2\,\rsol$ (D'Antona \& Mazzitelli, 1994) one finds that
 $\rslow=1.01\,\rstar$, 
$\ralfven=1.48\,\rstar$, and $\rfast=1.49\,\rstar$.

In summary, for any reasonable choice of TTS stellar and coronal
parameters, one 
finds that a thermally-driven stellar wind must be present\footnote{I
  do not propose that the observed 
outflows from young stellar objects are thermally-driven stellar
winds. These winds are much too weak to account for the inferred mass
outflow rates and mechanical luminosities. I have just shown that TTS
must have thermally-driven stellar winds, which probably coexist with
the more powerful, observed outflows, whose origin is still unknown.},
and an {\it upper\/} limit to the size of closed magnetic-structures
that may interact with the disk is  $r\la 3\,\rstar$. Therefore, 
there is no basis for assuming a closed field-geometry further out.
\section{Consequences for Magnetic Accretion}

Current models of magnetic-accretion rely on a closed stellar magnetic
field to mediate the accretion of disk matter and to spin down the
star, and a steady-state is assumed. In one way or another, 
the corotation  radius \rco---the 
 radius where the Keplerian angular velocity 
 equals the stellar angular velocity \omegastar---plays a
 fundamental role in these models, with
 $\rco=5\,\rstar (\mstar/0.5\,\msol)^{1/3}\,(\omegastar/10^{-5}\,{\rm
 s^{-1}})^{-2/3}\,(\rstar/2.5\,\rsol)^{-1}$. The results of the preceding
 section show that there is no basis for assuming a closed field
 geometry beyond $r\la 3\,\rstar\sim 0.5\,\rco$. 

Does this matter?

Ostriker \& Shu (1995) argue that the
stellar field will truncate the disk just inside of $\rco$, and the
field lines will be open just beyond \rco. In this ad hoc way, they
bypass the issue of field-shearing by the Keplerian flow. However, if the
stellar field-lines are radial beyond $r \la 3\,\rstar$,
disk-disruption may occur only inside of $\la 3\,\rstar$---if at all. 
Hence, the Ostriker \& Shu model is missing an
important piece of physics. This casts serious doubts on the validity
of their model.

The more physically-realistic model of Ghosh \& Lamb
(1978\markcite{gl78})  invoked by K\"{o}nigl (1991) 
relies on the balancing of the spin-up and spin-down torques from
inside and outside \rco, respectively, and includes
field-shearing due to the disk flow.
 In this model, the growth of the toroidal field, $B_\varphi$, is limited by
reconnection, and it is {\rm assumed} that a steady-state is
reached. However, note that the wind flow described in \S 3 becomes
trans-fast-magnetosonic beyond $\rfast=1.5$--$3\,\rstar = 0.3$--$0.7\,\rco$,
and, therefore, no torques can be
exerted on the star from beyond $r=0.3$--$0.7\,\rco$ as long as the
field lines are filled with a stellar wind.

This situation will last until  a flux-tube is filled
with disk matter all the way to the star. Because of the higher
density of disk-matter, the Alfv\'{e}n  speed along the flux-tube
will be lower than that when wind matter is present, and this {\it
  may\/} result in $\rfast < \rstar$; if this is the case,  then
disk-torques from beyond $\sim 0.5 \rco$ will act on the star. To
within order of 
magnitude, disk matter will fill a flux-tube all the way to the star
in a free-fall time, \tff.
However, a strong toroidal field  $B_\varphi \sim B_r$ will develop
on a timescale $\tshear\sim\left|\omegakepler-\omegastar\right|^{-1}$
($\sim \tff$ outside the neighborhood of \rco)
where \omegakepler\ is the local
Keplerian angular velocity in the disk. The presence of this toroidal
field has important consequences. 

First, a realistic stellar wind will not be confined to the equatorial
plane, but it will fill a large fraction of $4\pi$
steradians. Because the field lines carried by this wind are mostly
radial, only the field near the stellar equatorial plane
 will be strongly sheared by
the disk matter. Thus, a strong gradient in $B_\varphi^2$ will
exist parallel to the axis of rotation; and the flow---and the
field---will be deflected {\it away\/} from the disk on a timescale
$t_z\sim h\rho_d^{1/2}/B_\varphi$, where $h$ is
the disk scale-height at a distance $r$ from the star and
$\rho_d$ is a typical density in the disk {\it outer\/}
atmosphere at $r$. The disk matter will
start shearing the field as soon as  the field penetrates to a disk-depth
where $\rho_d\sim
B^2/\vkepler^2$, where \vkepler\ is the local Keplerian
speed. Therefore, $t_z\sim (h/r)\tff < \tff$, and {\it before\/} disk
matter fills a stellar flux tube all the way to the star,
 this flux tube will be disconnected from the disk
outer layers and the loading of disk matter onto the field will stop.
Eventually, as new matter flows from the stellar
surface dragging along magnetic field lines, the original geometry may be
restablished; but this shows that the assumption of steady-state
accretion on timescales $\sim \tff$ is unwarranted, and that the
transmission of torques to the star from beyond $r\sim 0.5\,\rco$ is
very ineffective. 

One may argue that, by averaging over several rotation periods, a net
torque will act on the star from beyond $\sim 0.5$ corotation radii.
However, the above analysis shows that 
disk-torques from a point at a distance $r$ from the star will last
for a fraction $h/r$ of the free-fall time---where $h$ is the disk's
local scale-height---before the magnetic field 
detaches from the disk; therefore, the duty-cycle of
the disk-torques from beyond $\sim 0.5\,\rco$ is likely to be very small. 

Second, strongly sheared fields are subject to 
 MHD instabilities (both ideal and resisitive), which  relieve the
 field-shear on a 
 timescale \tins---typically, $\tins \ll \tshear$---with a dramatic
 change in topology.

 Probably, the most relevant
 ideal-MHD instability in this problem is the Balbus-Hawley 
 instability (Balbus \& 
Hawley, 1991\markcite{bh91}). Though the growth-time for this
instability is of order  the shearing timescale, $\tshear$, the
 non-linear evolution of the instability results in a 
{\it turbulent\/} magnetic field for almost any initial field geometry
(see Stone et al., 1996\markcite{stal96}). Hence, on a timescale $\sim
\tshear$  the disk will be endowed with a
 magnetized corona that is not simply-connected  to the star.

As for resistive instabilities, 
astrophysically-relevant instabilities rely on resisitivity of
anomalous origin which becomes important after a finite
stress-threshold  is crossed (see, e.g., Priest, 1982). 
The outcome of these instabilities is usually
a {\it sudden\/} change of
topology---solar and stellar flares are prime examples---as opposed to a
steady-state configuration (see, e.g., Priest et. al,
1986\markcite{priest86}, and references therein; Forbes \&
Priest, 1984\markcite{fp84}). In the context of T-Tauri stars, the
simulations 
of Hayashi, Shibata, \& Matsumoto (1996\markcite{hsm96}) show 
that when anomalous resistivity with a finite stress-threshold is
included, the stellar field is sheared by the disk for a time $\sim\tshear$,
and thereafter the stresses are {\it dynamically\/} relieved ($\tins \ll
\tshear$) with a dramatic change in field-topology. Therefore, there
is no basis to assume that reconnection will result in a steady-state
topology that is known a priori.

In summary, the fact that the magnetic field will be mostly tangential
to the disk surface beyond $\sim 0.5\,\rco$ means that there is no
basis for the Ostriker \& Shu (1995) claim that the disk will be
truncated just inside \rco\ and that disk currents induced by a
penetrating field result in an open field-geometry
beyond this point.
 Furthermore, due to the strong vertical gradients in
$B_\varphi$ that will develop if the stellar field is mostly radial,
each interaction of the stellar field with the disk will last for less
than about a local rotation period, and the disk torques on the star from
beyond $r\sim 0.5\,\rco$ will be very ineffective because the field
will be deflected away from the disk in much less than a free-fall
time. This casts serious doubts on the applicability of the Ghosh \&
Lamb model to this problem. Finally, a realistic consideration
of MHD instabilities shows that there is no basis to assume a
steady-state field-topology that is known a priori.
\section{Conclusions}

I have critically examined the validity of
current magnetic-accretion models for classical T-Tauri stars in light
of the available data and our knowledge of magnetic fields in active,
low-mass dwarfs. I have shown that the  hot coronae around
TTS can sustain thermally-driven stellar winds, which change
the field topology into a radial configuration beyond $\la
3$ stellar radii. Therefore, there is no basis for assuming a closed field
topology further out.

Because the magnetic field will be mostly tangential
to the disk surface beyond $\sim 0.5$ corotation radii, the disk will
not be truncated beyond this point, and field shearing by the disk
has to be included in any realistic model.

The strong vertical gradients in the toroidal field that will develop
in the outer layers of the disk will limit each
interaction between the field and the disk to  about one
local Keplerian period, while the disk-torques from beyond $\sim
0.5\,\rco$ will be effective for only a small fraction of this time.
Therefore, the assumption of steady-state accretion
is not warranted, and the duty-cycle of disk torques on the star from
beyond about half the corotation radius will be small.

 Finally, a realistic consideration
of MHD instabilities shows that there is no basis to assume a
steady-state field-topology that is known a priori.

How, then, does non-equatorial accretion take place?

It is beyond the scope of this paper to present a fully-consistent
model. On the contrary, it is my purpose to point out that much work needs to
be done before such model can be presented. In particular, we need a
better understanding of the interaction of the stellar field with the
disk atmosphere through MHD instabilites, and the inherent
time-dependent nature of the problem cannot be ignored.

Magnetic accretion probably {\it does\/} take place, but at distances
smaller than those proposed in current models.
I have shown that the
stellar magnetic field is likely to be radial {\it beyond\/} $\la 3$
stellar radii; inside this radius, magnetic accretion may be mediated
by stellar magnetic loops. However, because these magnetic loops only
exist well inside the corotation radius, the accreting material will
spin-up the star rather than spin it down (see Wang 1997\markcite{wang97}).
Therefore, the observational evidence for classical T-Tauri stars being
slower rotators than their naked siblings is still unexplained.

\acknowledgments

I am grateful to Eugene Parker for a very encouraging afternoon
discussion at the University of Chicago, and to John Contopoulos for
his points on the transmission of signals in MHD flows and his
critical reading of a first draft. I also wish to thank Steve Martin,
Lee Mundy, Steve Stahler, Stephen White, and Ellen Zweibel for their
critical reading of the manuscript and many useful comments. I am also
indebted to an anonymous referee for
comments and criticisms which significantly improved the
original version of this paper.

This work was  supported by NSF grant AST9314847 to the Laboratory
for Millimeter-Wave Astronomy at the University of Maryland.

\clearpage
%
%

\figcaption{Sketch of the magnetic topology in the corona of a T-Tauri
  star with a thermally-driven wind. Two magnetic elements of opposite
  polarity are shown (not to scale), and the 
  {\it hatched\/} regions correspond to magnetic flux tubes with an
  enhanced density; these flux tubes would appear as X-ray-bright
  coronal loops. For clarity, the magnetic topology in the
  photosphere, chromosphere, and transition region are not shown. The
  {\it arrowheads\/} indicate the polarity of the field lines.
}
%
%

\figcaption{Radius where the thermal pressure dominates over magnetic
  stresses, \ropen\, (in units of the stellar radius \rstar), as a
  function of the mean dipole separation at the stellar surface, $d$
  (also in units of \rstar), for a T-Tauri star of mass
  $\mstar=0.5\,\msol$, radius $\rstar=2.5\,\rsol$, and coronal
  temperature $\tstar=10^6\,\kelvin$. Each {\it hatched \/} region
  corresponds 
  to one value of the 
  magnetic field strength at the base of the corona, $B_\ast$, and to
  coronal-base densities, $n_\ast$, in the range
  $10^8$--$10^9\,\pccm$. The {\it 
  upper\/} and {\it lower \/} curves that bound each region correspond to
  $\ropen(d)$ for $n_\ast =
  10^8\,\pccm$ and $n_\ast =  10^9\,\pccm$, respectively; the labels
  indicate the value of $B_\ast$ in Gauss. 
}
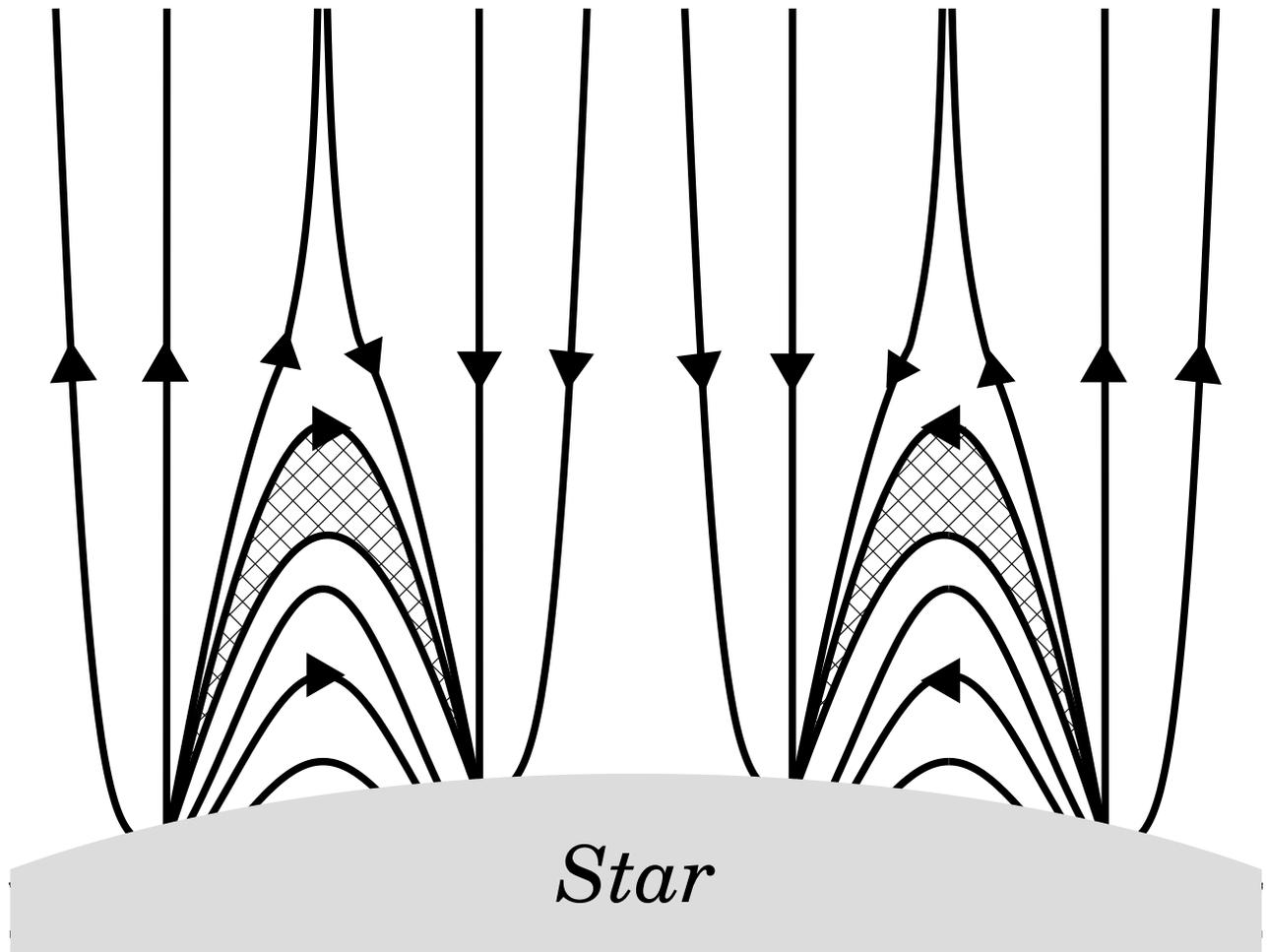
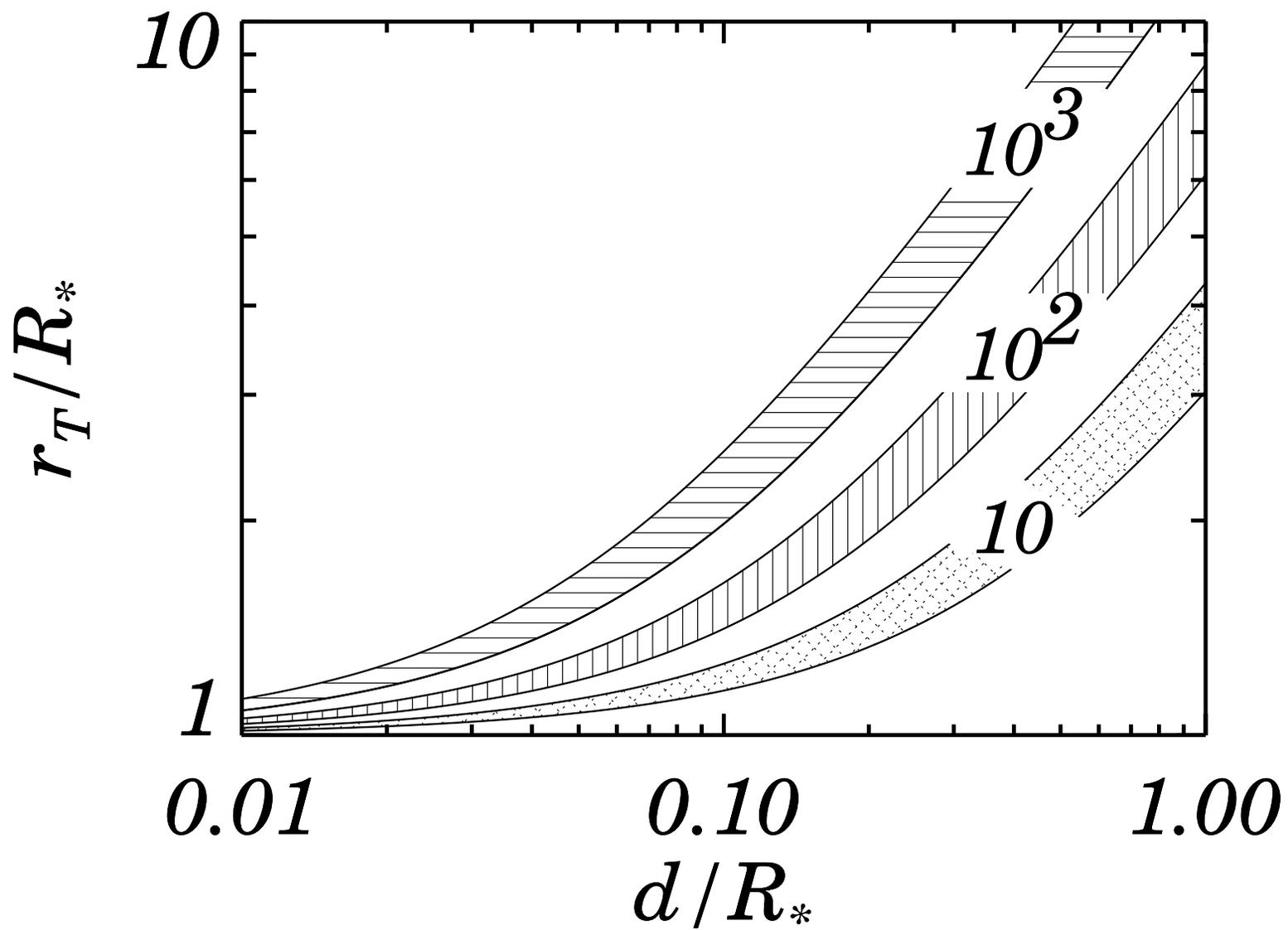
\end{document}